\documentclass[twocolumn,times,tighten,10pt]{aastex631} 
\usepackage{amsmath,amstext,mathrsfs}
\usepackage{epsfig}
\usepackage{epstopdf}
\usepackage{color} 
\usepackage{natbib}
\usepackage{gensymb}
\usepackage{natbib}
\usepackage{amssymb}
\usepackage{bm}
\usepackage{rotating}
\usepackage{graphicx}
\def\apjl{{ApJ}}%
\def\apjs{{ApJS}}%
\def\aap{{A\&A}}%
\shorttitle{Magnetic fields in the vicinity of Monogem TeV Halo}
\shortauthors{Malik et al.}

\begin{document}
\title{Study of Magnetic Field and Turbulence in the TeV halo around Monogem Pulsar}

\author[0000-0003-4147-626X]{Sunil Malik}
\affiliation{Institute fur Physik und Astronomie Universitat Potsdam, Golm Haus 28, D-14476 Potsdam, Germany}
\affiliation{Deutsches Elektronen-Synchrotron DESY, Platanenallee 6, 15738 Zeuthen, Germany}

\author[0000-0003-1683-9153]{Ka Ho Yuen}
\affiliation{Theoretical Division, Los Alamos National Laboratory, Los Alamos, NM 87545, USA}

\author[0000-0003-2560-8066]{Huirong Yan}
\affiliation{Institute fur Physik und Astronomie Universitat Potsdam, Golm Haus 28, D-14476 Potsdam, Germany}
\affiliation{Deutsches Elektronen-Synchrotron DESY, Platanenallee 6, 15738 Zeuthen, Germany}
\email{huirong.yan@desy.de}

\begin{abstract}

Magnetic fields are ubiquitous in the interstellar medium, including extended objects such as supernova remnants and diffuse halos around Pulsars. Its turbulent characteristics govern the diffusion of cosmic rays and the multi-wavelength emission from PWNe. However, the geometry and turbulence nature of the magnetic fields in the ambient region of PWN is still unknown. Recent gamma-ray observations from HAWC and synchrotron observations suggest a highly suppressed diffusion coefficient compared to the mean interstellar value. In this study, we present the first direct observational evidence of the orientation of the mean magnetic field and turbulent characteristics by employing a recently developed statistical parameter `Y$_{turb}$' in the extended halo around the Monogem pulsar. Our study points two possible scenarios: nearly aligned toward the line of sight (LoS) with compressible modes dominance or high inclination angle toward LoS and characterized by Alfv\'enic turbulence. The first scenario appears consistent with other observational signatures. Furthermore, we report that the magnetic field has an observed correlation length of approximately $3 \pm 0.6 {\rm pc}$ in the Monogem halo. Our study highlights the pivotal role of magnetic field and turbulence in unraveling the physical processes in TeV halos and cosmic ray transport.

\end{abstract}


\section{Introduction}
\label{introduction}

The interstellar medium (ISM) is permeated by the magnetic fields, spanning a wide range of strengths from $\mu {\rm G}$ to mG~\citep{Reynolds2012SSR}. The turbulent characteristics of this magnetic field play a crucial role in facilitating various physical processes, including particle transport and acceleration~\citep{schlickeiser2002statistical,2002PhRvL..89B1102Y, YLP2008, Lemoine22, Yan2022rev}. Magnetohydrodynamic (MHD) turbulence is generally classified into the Alfvénic, Compressible fast and slow modes ~\citep{CL03, LP12, wang2020,Makwana2020,leakage}. The compressible modes interacts significantly with cosmic rays, governing their scattering and diffusion across different phases of the ISM as well as in the ambient regions of extended objects~\citep{YL2004ApJ, Yan2008ApJ, Lynn2013}. Therefore, to comprehensively understand cosmic ray diffusion and multi-wavelength emissions in the extended halo around pulsars, it is essential to investigate the underlying magnetic field configuration and the nature of the associated turbulence.

Recent studies~\citep{Abeysekara2017Sci, Cristofari2021A&A} reported that some of the pulsars host the exotic extended TeV and PeV halos around them. In particular, it was discovered that the diffusion in the Geminga and Monogem halo is two orders of magnitude lower than the averaged value in ISM ~\citep{Abeysekara2017Sci, Martin2022A&A}.
It was suggested that the suppressed diffusion in the extended halo around the pulsar can be attributed to the topology of the local magnetic fields~\citep{Liu2019PRL}. There are also other mechanisms proposed ~\cite[e.g.][]{Fang2019MNRAS, Recchia2021PhRvD}. The detection of local magnetic field and turbulence holds the key to distinguish the different scenarios.

However, the direct observation of the orientation and strength of the magnetic field in the ambient halo region of pulsars has proven to be challenging. To address this issue, our recent study has employed turbulence mapping theory based on synchrotron emission statistics, presenting a novel statistical approach using the parameter called {\rm $Y_{turb}$}~\citep{Malik2023}. This technique has the potential to unveil the 3D geometry of magnetic fields and their dominant MHD turbulence characteristics. In this paper, we utilize the {\rm $Y_{turb}$} to provide the first observational investigation to constrain the magnetic field direction within these halos. 

This paper is structured as follows: In Section \ref{pwn}, we examine the characteristics of the TeV halo, with particular emphasis on the extended halo of the Monogem pulsar. Our primary analysis of diagnosing the mean magnetic field inclination angle, utilizing the recently developed turbulence anisotropy analysis and the multi-points structure function, is described in Section \ref{sec:obs}. The discussions and summary of our findings are presented in Sections \ref{discuss} and \ref{sec:conclusion}, respectively.

\section{Characteristic of Extended TeV halo around pulsars}
\label{pwn}

The HAWC collaboration recently made notable discoveries of potential TeV sources, namely 2HWC J0635+180 and 2HWC J0700+143~\citep{2017ApJ...843...39A}, in the vicinity of Geminga and Monogem at significance levels of 13.1$\sigma$ and 8.1$\sigma$ respectively~\citep{2017ApJ...843...39A, Abeysekara2017Sci}. These sources are located at a relatively short distance from Earth, with Geminga at $250 \ {\rm pc}$ and Monogem at $288 \ {\rm pc}$. The diffusive propagation of particles within these halos extends to approximately $10-50 \ {\rm pc}$, which points to a much slower diffusion coefficient compared to the Galactic mean, presenting a challenge in comprehending the underlying mechanisms responsible for the observed high-energy emissions~\citep{Abeysekara2017Sci}.

In~\cite{Liu2019PRL}, the authors proposed a promising model for explaining the TeV observations of the extended halo of the Geminga using anisotropic diffusion. In the case of sub-Alfv\'enic turbulence, the diffusion coefficients perpendicular and parallel to the mean magnetic field show significant differences (i.e., $D_{\perp} \sim D_{||} M_{A}^{4}$) (where $D_{\perp}$, $D_{\parallel}$ is diffusion coefficient in perpendicular and parallel direction of the magnetic field, respectively, $M_A$ represents Alfv\'enic Mach Number)~\citep{Yan2008ApJ, Giacinti2012PhRvL,lopez2018MNRAS, Maiti2022ApJ}. The authors of this theoretical model inferred that local magnetic fields with small inclination angles relative to the line of sight could account for the observed TeV emissions and the phenomenon of suppressed diffusion. Currently, direct measurements of the magnetic field strength and orientation are not feasible due to the absence of synchrotron polarized emissions in the Geminga halo.


To get the first direct observational evidence, in this study, we focus on extended halo of the Monogem which is observed in multi-wavelength from radio to gamma rays~\citep{Uyaniker1999A, Abeysekara2017Sci}. There is a pulsar (PSR B0656+14) inside the extended halo ~\citep{Johnston2006MNRAS,john2007MNRAS}. The pulsar has a period (P) of approximately $385 \ {\rm ms}$, period derivative ($\dot{\rm P}\sim 5.4 \times 10^{-14} \ {\rm s}\ {\rm s}^{-1}$) corresponding\footnote{See e.g. ~\cite{Lorimer2004book} for the formulae used to derive $\rm{\tau}$, $\dot{{\rm E}}_{rot}$ and $\rm{{\rm B}_{surf}}$ using ${\rm P}$ and ${\rm \dot{\rm P}}$} to a spin-down age ($\tau$) of approximately $111 \ {\rm kyr}$, a spin-down power (${\rm E}_{rot}$) of $3.8 \times 10^{34}$ erg ${\rm s}^{-1}$, and a surface magnetic field (${\rm B}_{surf}$) of approximately $4.65 \times 10^{12}$ G.

\section{3D magnetic field from turbulence analysis}
\label{sec:obs}
\subsection{Synchrotron Polarisation Observations}

In order to investigate the strength and morphology of magnetic fields in the Monogem TeV halo region, we utilized the {\tt Effelsberg 100-m 21 cm POL}\footnote{https://www3.mpifr-bonn.mpg.de/survey.html} survey. This survey, conducted at medium Galactic latitudes, involves radio continuum and polarization observations. The data were obtained using the Effelsberg 100-m telescope at a frequency of $1.4 \ {\rm GHz}$, with an angular resolution of $9.35^{'}$ ~\citep{Uyaniker1999A}.

\begin{figure}
    \centering
    \includegraphics[width=0.48\textwidth]{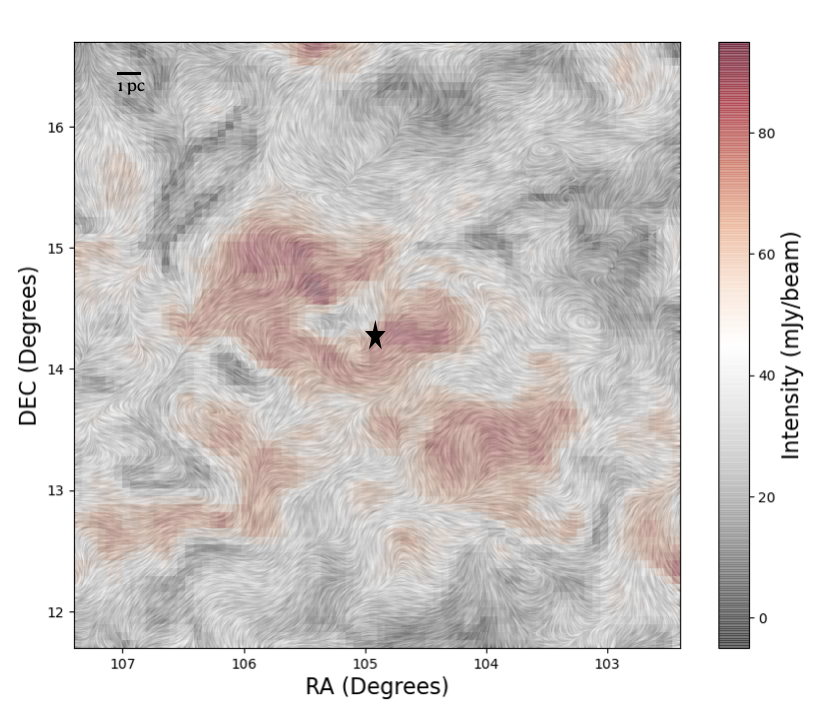}
    \caption{The synchrotron polarisation emission map of extended halo of the Monogem ($5\degree\times5\degree$) with the pulsar location indicated by star symbol at ${\rm R.A.(J2000)} = {\rm 06^{h}59^{m}48^{s}.02} \ (104.94^{\circ}), {\rm Dec.(J2000)} = 14\degree14^{'}19^{''}.4 \ (14.23^{\circ})$ using Effelsberg 100-m telescope observation at $1.4 \ {\rm GHz}$ from ~\cite{Uyaniker1999A}. Here the color bar represents the flux in mJy/beam. The streamline, computed via the Line Integral Convolution (LIC) algorithm~(\citealt{Leedom}, \citealt{ZOCKLER1997975}),  represents the direction of the magnetic field inferred from the synchrotron polarisation observations.} 
    \label{synchro}
\end{figure}

We specifically employed $5\degree\times5\degree$ synchrotron polarization observations of the extended halo of the Monogem, as depicted in Fig.~\ref{synchro}. The observation map was overlaid with streamlines computed using the Line Integral Convolution (LIC) algorithm, which indicates the inferred direction of the magnetic field based on the polarization direction~\citep{Leedom, ZOCKLER1997975}. The figure exhibits a significant increase in linearly polarized flux in and around the vicinity of the Monogem Pulsar, which is indicated by the black star symbol. The polarisation streamlines clearly show the variations in the map. The spherical dispersion of the observed polarisation angle ($\chi$) in the region suggests a 2D Alfv\'en Mach number in the pictorial plane of sky is approximately $M_{A,2D}\approx 1.0$. Utilizing this synchrotron map, we investigated the magnetic field topology in the extended halo of Monogem.

\subsection{Correlation length from Structure-Function Analysis}
\label{result2}

\begin{figure}
\centering
    \includegraphics[width=1.0\columnwidth]{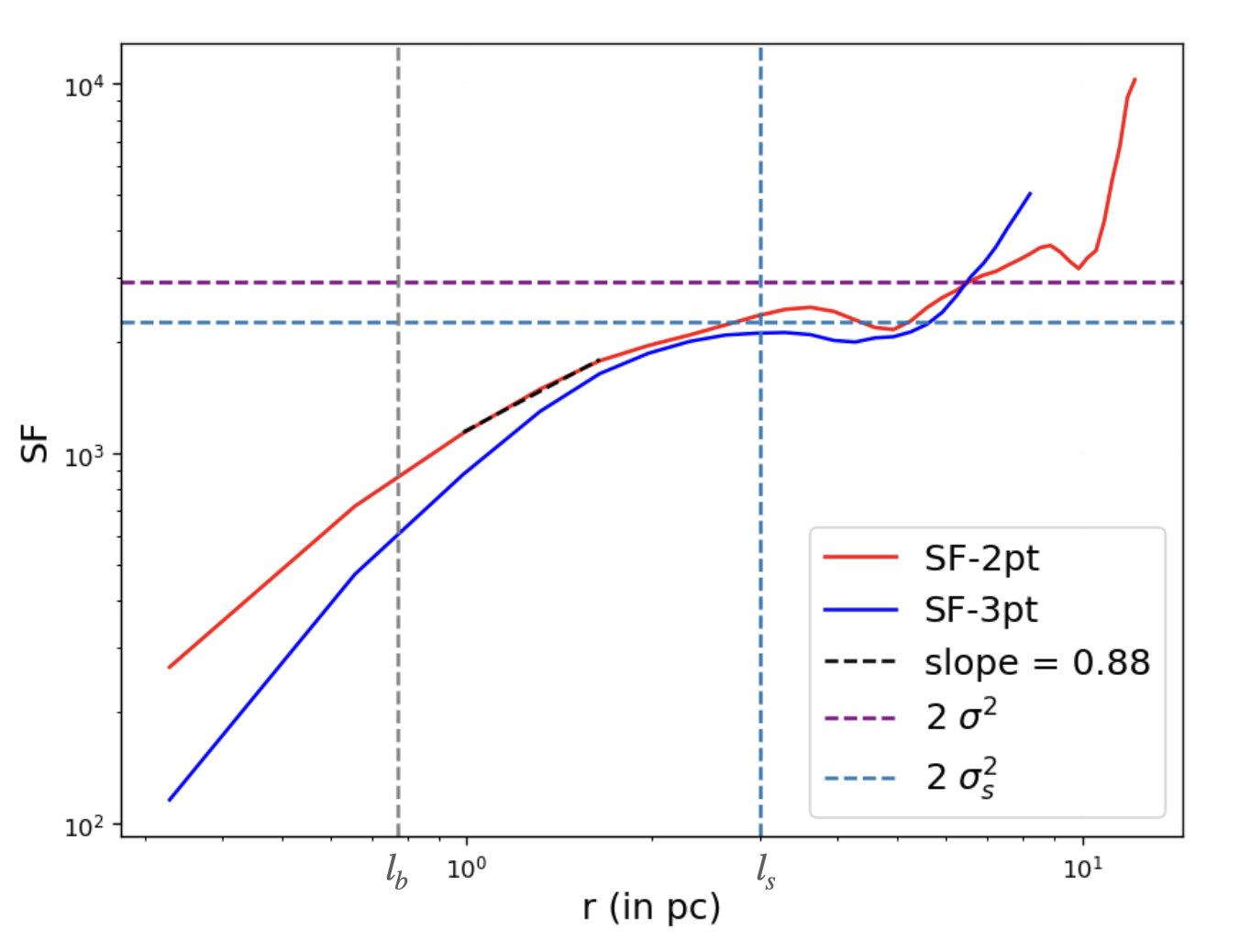}
    \caption{The multi-point structure function for the total polarised intensity. Different colors correspond to different points' structure functions estimated using Eq.~\ref{sf2} \&~\ref{sf3}. The grey and blue dashed vertical lines indicate the telescope's beam size $l_b$ and $l_s$, respectively. The purple and blue dashed horizontal lines mark $2 \sigma^{2}$ and  $2 \sigma_{s}^{2}$, corresponding to the variance for initial polarised intensity and small-scale fluctuations, respectively. The black dashed line indicates the fitting line having slope $\sim0.88$.}
    \label{SF}
\end{figure}


In polarisation observations, the large-scale magnetic fields can coexist with the small-scale turbulent components. In recent studies~\citep{hildebrand2009ApJ,pattle2017ApJ,cho2019ApJ}
several approaches have been discussed to address this issue, including the fitting method and wavelet transformation~\citep{Grebenev1995ApJ}. These techniques heavily rely on data coarsening. One widely accepted and commonly used method for assessing fluctuations is through two-point structure function (SF), which has been discussed in multiple literatures on potential physical parameter retrieval \citep{LP12, KLP16,2016ApJ...831...77L,2018ApJ...855...29H,2019ApJ...877..108L,2019ApJ...887..258H,leakage}, defined as follows:

\begin{equation}
    {\rm SF^{2pt}_{2}}({\bf R}) = \langle|X({\bf R}'+{\bf R})-X({\bf R})|^ 2\rangle_{{\bf R}'}
    \label{sf2}
\end{equation}
where $X$ represents an observable in the plane of the sky. ${\rm SF^{2pt}_{2}}$ will be an increasing function of ${\bf R}$ when we are in the regime of ${\bf R}< l_s$, here $l_{s}$ represents the correlation length of small-scale fluctuations. It will attain a plateau for ${\bf R}\gtrsim l_s$. However, the shape of ${\rm SF^{2pt}_{2}}$ at large ${\bf R}$ gets distorted in the presence of large-scale shearing in the observables. In this case, three and more point second-order structure functions can be utilized to explore these large-scale effects ~\citep{cho2019ApJ}. The three points SF is expressed as;
\begin{equation}
    {\rm SF^{3pt}_{2}}({\bf R}) = \frac{1}{3}\langle|X({\bf R}'-{\bf R})-2 X({\bf R}')+X({\bf R}'+{\bf R})|^ 2\rangle_{{\bf R}'}.
    \label{sf3}
\end{equation}

The two-point structure function is represented by the red line in Fig~\ref{SF}. From the plot, it is evident that our region of Monogem extended halo exhibits both large-scale structures and small-scale fluctuations in the magnetic field. The three points ($3 \ {\rm pts}$) SF is depicted by the blue line in Fig.\ref{SF}. Structure functions with more than two points filter out the influence of large-scale magnetic field structures on determining the correlation length of small-scale turbulent fields, as evident in our SF plot in Fig.\ref{SF}. The two horizontal line corresponding to $2 \sigma^{2}$ and $2 \sigma_{s}^{2}$, where $\sigma$ and $\sigma_{s}$ represent, respectively, the standard deviation of the total fluctuations and the small-scale turbulent fluctuations after removal of large scale fluctuations using the annular average technique from ~\citep{cho2019ApJ}. The ${\rm Y_{turb}}$ parameter is calculated then based on only the small-scale turbulent fluctuations.

Consequently, we determined that the region possesses a characteristic observed correlation length of approximately $3\ {\rm pc}$ for the MHD turbulence, based on the synchrotron map. The correlation length, $l_{s} \sim 3 \pm0.6\ {\rm pc}$, serves as the upper bound to conduct the turbulence anisotropy analysis as described in \S\ref{Y_ana}.

\begin{figure*}
\centering
    \includegraphics[scale=0.65]{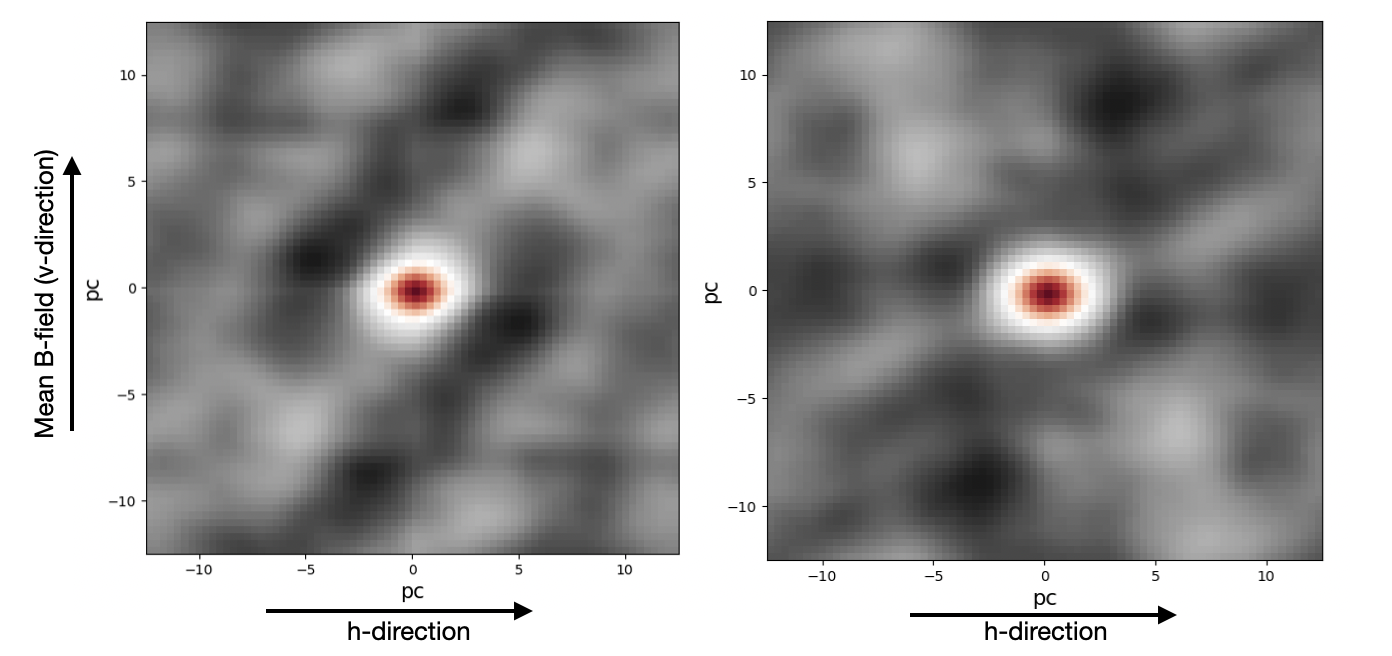}
    \caption{ {\it Left panel:} This figure shows the correlation function distribution for the $I+Q \propto B_{x}^{2}$. {\it Right panel:} It is same as the left but for $I-Q\propto B_{y}^{2}$. Here we have marked the v-direction and h-direction along the y and x-axis, respectively. The direction of the plane of sky component of the magnetic field, $B_{\perp}$ is in the v-direction.}
    \label{anisotropy}
\end{figure*}

\subsection{Description of {\rm $Y_{turb}$}-parameter technique}
\label{Y_tech}

To measure the relative angle between the mean magnetic field and the line of sight, we employ a recently developed statistical technique based on synchrotron polarisation map~\citep{Malik2023}. It was suggested in earlier literature that the anisotropy of the Stokes parameters can be used to infer the geometric properties of the magnetic field \citep{leakage}. The global correlation function, represented as $D_X$, can effectively capture the anisotropy of a given observable, denoted as $X$, which can be expressed as follows:

\begin{equation}
    D_X({\bf R}) = \langle (X({\bf R}') X({\bf R}'+{\bf R}))^2\rangle_{{\bf R}'}.
    \label{eqd}
\end{equation}

\cite{Malik2023} suggest that the {\rm $Y_{turb}$}, which is expressed as:

\begin{equation}
Y_{turb} = \frac{\text{Anisotropy}(D_{I+Q})}{\text{Anisotropy}(D_{I-Q})} = \frac{R_{v}/R_{h}(D_{I+Q})}{R_{v}/R_{h}(D_{I-Q})}.
\label{eq_y}
\end{equation}

 can be used to trace the inclination angle of the magnetic field.  Here, R$_{v}$ and R$_{h}$ represent the extent of correlation function distribution in the vertical and horizontal direction as shown in Fig.\ref{anisotropy}, whereas  B$_{\perp}$ direction is toward the vertical axis. Here the plane of sky B$_{\perp}$ direction is determined by mean polarisation angle of the map. This statistical parameter, derived from polarized synchrotron radiation, captures the variations in the embedded magnetic fields resulting from turbulence. Consequently, it enables us to investigate the geometry of the 3D magnetic field and the characteristics of the underlying MHD turbulence. 
 
In ~\cite{Malik2023}, we computed the ${\rm Y_{turb}}$ parameter for both solenoidal and compressible simulated MHD cubes. 
 With the analysis of total and decomposed (Alfv\'enic (A) mode and Compressible (C) Mode) datacubes, 
 we found that the relative anisotropy of Stokes parameters can act as a diagnostic for retrieving the magnetic field inclination and identifying the dominating MHD turbulence mode in ISM and extended sources like Monogem TeV halo. This anisotropy has been quantified by taking the Gaussian extents obtained using 2D Gaussian fitting of the correlation function distribution in v and h-directions. We established a statistical criterion of ${\rm Y_{turb}} \sim 1.5\pm0.5$ (with ${\rm Y_{turb}} > 1.5$ indicating the A-mode and ${\rm Y_{turb}} < 1.5$ indicating the C-mode) to identify the dominant fraction of MHD turbulence modes.  
 Remarkably, the Y-parameter for different modes exhibited contrasting trends (either decreasing or increasing) with respect to the mean magnetic field inclination angle with respect to the line of sight to the observer, denoted as $\theta_\lambda$.

\subsection{Line of sight inclination angle of B-field from {\rm $Y_{turb}$} Analysis}
\label{Y_ana}

\begin{table}
\centering
\begin{tabular}{ cccc} 
 \hline
 \hline
  r (pc) & $R_{v}$/$R_{h}$($D_{I+Q}$) & $R_{v}$/$R_{h}$($D_{I-Q}$) & ${\rm Y_{turb}}$ \\ 
 \hline
 0.98 & 0.95 & 1.01 & 0.94 \\ 
 \hline
 1.31 & 0.91 & 0.97 & 0.93 \\
 \hline
 1.64 & 0.87 & 0.93 & 0.92 \\ 
 \hline
 1.97 & 0.83 & 0.90 & 0.92 \\ 
 \hline
 2.30 & 0.81 & 0.87 & 0.92 \\
 \hline
 \hline
\end{tabular}
\caption{The table shows the possible values of ${\rm Y_{turb}}$ estimated using different fitting radii to capture anisotropy of correlation function distributions for (I+Q) and (I-Q).}
\label{ytable}
\end{table}

To estimate the mean-field inclination angle for the vicinity of the Monogem TeV halo, we applied the technique described earlier to synchrotron polarization observations (Fig.\ref{synchro}) within a ($5^\circ \times 5^\circ$) region around the Monogem pulsar. We initially analyzed the correlation function of $I+Q \propto B_x^2$ and $I -Q \propto B_y^2$ and plotted the results in Fig.~\ref{anisotropy}. These correlation function plots clearly depict the difference in horizontal and vertical lengths relative to the mean magnetic field direction. To quantify the anisotropies, we performed 2D Gaussian fittings on these distributions using different fitting radii within the observed correlation length of $3.0$ pc and obtained ${\rm Y_{turb}} < 0.94$. We have summarised possible values of ${\rm Y_{turb}}$ in Table~\ref{ytable}. According to our previous investigation in ~\cite{Malik2023}, for ${\rm Y_{turb}} < 1.0$, we have identified two potential scenarios: one with $\theta_\lambda < 10^\circ$ dominated by compressible modes, and the other with $\theta_\lambda > 60^\circ$ characterized by Alfv\'enic turbulence. The first scenario, i.e., the small inclination angle of the mean magnetic field is consistent with such as gamma-ray measurement of the halo from HAWC \citep{Abeysekara2017Sci} and no detection of X-ray though we can not rule out the second scenario of Alfven dominance with a large inclination angle. 
The small coherence length seems consistent with \cite{lopez2018MNRAS}, though it is still unclear that how the suggested enhanced isotropic Kraichnan/Kolmogorov turbulence on small scale can be generated.

These results clearly indicate that the extended halo exhibits a mean magnetic field that is potentially aligned with the line of sight, along with the presence of compressible turbulence. This alignment is crucial for the propagation of cosmic rays~\citep{Yan2008ApJ}. Importantly, our analysis provides direct observational confirmation for the speculation that the magnetic fields in such TeV halos are not in the extreme limit, as suggested in the existing literature. This finding offers reassurance regarding the effectiveness of our statistical technique using {\rm $Y_{turb}$} in retrieving 3D magnetic fields in various environments, including the ISM and extended halo around pulsars.

\label{sec:analysis}

\section{Discussion} 
\label{discuss}

As discussed in Section~\ref{introduction}, MHD turbulence plays a crucial role in the acceleration and scattering of cosmic rays in the ISM and other extended objects \cite[see also][]{Yan2012ApJCR}. The recent discovery of TeV halos around middle-aged pulsars, such as Geminga and Monogem (PSR B0656+14), has raised intriguing questions, particularly the origin of the slower diffusion~\citep{Abeysekara2017Sci}. In a recent study by \cite{Liu2019PRL}, it was proposed that sub-Alfvenic turbulence combined with a local mean magnetic field aligned with the line of sight can account for the missing X-ray flux as well as the slow diffusion since $D_{\perp} \simeq D_{\parallel} M_{A}^{4}$ according to \cite{Yan2008ApJ}. Therefore, to obtain a comprehensive understanding of these observations, it is crucial to directly investigate the 3D magnetic field strength, orientation, and the underlying MHD turbulence.

To investigate the magnetic field topology in extended objects such as TeV halos associated with pulsars, direct observational evidence is currently lacking. A recent study by \cite{Liu2019ApJ} has suggested an upper limit of $0.8 \ \mu {\rm G}$ for the magnetic field in the $10^{'}$ area surrounding the Geminga pulsar, which is lower than the average galactic magnetic field strength ($\sim 5.5 \ \mu G$, \citealt{Sofue2019MNRAS}). The absence of radio-polarized emission in the vicinity of Geminga has hindered direct observations of the magnetic field strength and associated turbulence in that region.

Furthermore, synchrotron polarised radiation encounters foreground Faraday rotation ($\theta_{FR}$) from the foreground, which may influence this technique’s applicability. From the synthetic observations through the turbulent foreground, we find that the Faraday rotation has a large impact on ${\rm Y_{turb}}$ in particular to the ambiguous range of $1<{\rm Y_{turb}}<2$. Hence, it is highly probable that the low ${\rm Y_{turb}}$ value $0.92$ we detected points to a low inclination angle of the magnetic field. In our case, the only available rotation measure is that of the Monogem pulsar itself, about $23.0\pm3.0 \ {\rm rad \ m^{-2}}$ from~\cite{john2007MNRAS}, suggesting a small likelihood of existence of clumpy structure along the line of sight. To fully assess the impact of foreground Faraday rotation, future Faraday tomography \citep[see, e.g.,][]{Takahashi2023PASJ} with multi-wavelength polarisation observations will be necessary.

Furthermore, in the Monogem TeV halo, the required diffusion coefficient is also significantly suppressed compared to the ISM value, similar to the case of the Geminga TeV halo. This suppressed diffusion, combined with the small $\theta_\lambda$ angle, suggests that cross-field transport with $D_\perp\propto D_\| M_A^4$ serves as the underlying mechanism for the reduced diffusion. We have considered $D_\|\simeq D_{ISM}\sim 3.8\times 10^{28} (E_{e}/1 {\rm GeV})^{1/3} \ {\rm cm}^2 \ {\rm s}^{-1}$, the average ISM diffusion coefficient as adopted in ~\cite{Liu2019PRL}. Assuming $D_{obs}\sim D_\perp \sim 4.5 \times 10^{27} \ {\rm cm}^2 \ {\rm s}^{-1}$ for 100 TeV particles as reported by ~\cite{Abeysekara2017Sci}, we estimate a local Alfv\'enic Mach number $M_A\sim 0.22$. The results obtained using various diagnostics are summarized in the table~\ref{table1}. 

\vspace{0.2cm}

\begin{table}[h]
\centering
\begin{tabular}{ ccc} 
 \hline
\hline
Diagnostics & Measure & Values  \\ \hline \hline
$Y_{\rm turb}$-parameter  & $\theta_{\lambda}$&$<10^\circ$\\ \hline
Structure Function&$l_{s}$& $\sim 3.0$ pc\\ \hline
Polarisation Angle &  $M_{A,2D}$ & 1.0 \\ \hline
${D_\perp/D_\|}$  &  $M_{A,3D}$ & $\gtrsim 0.2$ \\  
\hline \hline
\end{tabular}
\caption{We have listed all the estimated quantities in this study.}
\label{table1}
\end{table}
\vspace{0.5cm}


Additionally, in the halo-like environment, there are limited methods to gain insight into the strength of the parallel magnetic field component ($B_\|$). The rotation and dispersion measures of the pulsar might offer some clues about the $B_\|$ component\footnote{Using RM of $23.0\pm3.0 \ {\rm rad} \ {\rm m}^{-2}$ and DM of $13.7\pm0.2$ pc cm$^{-3}$, we can estimate the $B_\| \sim 2.06\pm0.72 \mu$G.} of the magnetic field, but are also affected by foreground contamination. In any case, the magnetic field orientation can not be inferred from the observations of pulsars themselves. Therefore, our study facilitates a pivotal advancement in the understanding of magnetism and the underlying physical process governing the TeV-PeV halos associated with pulsars.

\section{Conclusions}
\label{sec:conclusion}

In this study, we have presented the first observation analysis of magnetic field 3D geometry and underlying MHD turbulence nature in the extended TeV halo associated with the Monogem pulsar using radio polarisation observations. Our structure-function analysis shows that the region has a pictorial plane correlation length of $\sim 3\pm0.6\ {\rm pc}$ for the MHD turbulence in the region. Further, this analysis using synchrotron polarisation observations with ${\rm Y_{turb}} \simeq 0.92$ suggests two potential scenarios: one with $\theta_\lambda < 10^\circ$ and is dominated by compressible turbulence fluctuations, and the other with $\theta_\lambda > 60^\circ$ characterized by Alfv\'enic turbulence. Particularly, the first scenario is in line with other observational signatures (see table~\ref{table1}), and can account for the suppressed diffusion of cosmic rays as observed by HAWC~\citep{Abeysekara2017Sci}. 

Our study underpins the crucial role of the magnetic field in understanding the physical processes involved in recent and future detections of such extraordinary high-energy emissions in the Milky Way.\\

\section*{Acknowledgments}
We thank the anonymous referee for valuable comments that significantly improved the paper. SM would like to thank Parth Pavaskar, Siqi Zhao, and Bingqiang Qiao for the helpful discussions. The research presented in this article was partially supported by the Laboratory Directed Research and Development program of Los Alamos National Laboratory under project number(s) 20220700PRD1.

\bibliography{references}

\begin{thebibliography}{}
\expandafter\ifx\csname natexlab\endcsname\relax\def\natexlab#1{#1}\fi
\providecommand{\url}[1]{\href{#1}{#1}}
\providecommand{\dodoi}[1]{doi:~\href{http://doi.org/#1}{\nolinkurl{#1}}}
\providecommand{\doeprint}[1]{\href{http://ascl.net/#1}{\nolinkurl{http://ascl.net/#1}}}
\providecommand{\doarXiv}[1]{\href{https://arxiv.org/abs/#1}{\nolinkurl{https://arxiv.org/abs/#1}}}

\bibitem[{{Abeysekara} {et~al.}(2017{\natexlab{a}}){Abeysekara}, {Albert},
  {Alfaro}, {Alvarez}, {{\'A}lvarez}, {Arceo}, {Arteaga-Vel{\'a}zquez}, {Avila
  Rojas}, {Ayala Solares}, {Barber}, {Bautista-Elivar}, {Becerril},
  {Belmont-Moreno}, {BenZvi}, {Berley}, {Bernal}, {Braun}, {Brisbois},
  {Caballero-Mora}, {Capistr{\'a}n}, {Carrami{\~n}ana}, {Casanova}, {Castillo},
  {Cotti}, {Cotzomi}, {Couti{\~n}o de Le{\'o}n}, {De Le{\'o}n}, {De la Fuente},
  {Dingus}, {DuVernois}, {D{\'\i}az-V{\'e}lez}, {Ellsworth}, {Engel},
  {Enr{\'\i}quez-Rivera}, {Fiorino}, {Fraija}, {Garc{\'\i}a-Gonz{\'a}lez},
  {Garfias}, {Gerhardt}, {Gonz{\'a}lez Mu{\~n}oz}, {Gonz{\'a}lez}, {Goodman},
  {Hampel-Arias}, {Harding}, {Hern{\'a}ndez}, {Hern{\'a}ndez-Almada}, {Hinton},
  {Hona}, {Hui}, {H{\"u}ntemeyer}, {Iriarte}, {Jardin-Blicq}, {Joshi},
  {Kaufmann}, {Kieda}, {Lara}, {Lauer}, {Lee}, {Lennarz}, {Vargas},
  {Linnemann}, {Longinotti}, {Luis Raya}, {Luna-Garc{\'\i}a}, {L{\'o}pez-Coto},
  {Malone}, {Marinelli}, {Martinez}, {Martinez-Castellanos},
  {Mart{\'\i}nez-Castro}, {Mart{\'\i}nez-Huerta}, {Matthews},
  {Miranda-Romagnoli}, {Moreno}, {Mostaf{\'a}}, {Nellen}, {Newbold}, {Nisa},
  {Noriega-Papaqui}, {Pelayo}, {Pretz}, {P{\'e}rez-P{\'e}rez}, {Ren}, {Rho},
  {Rivi{\`e}re}, {Rosa-Gonz{\'a}lez}, {Rosenberg}, {Ruiz-Velasco}, {Salazar},
  {Salesa Greus}, {Sandoval}, {Schneider}, {Schoorlemmer}, {Sinnis}, {Smith},
  {Springer}, {Surajbali}, {Taboada}, {Tibolla}, {Tollefson}, {Torres},
  {Ukwatta}, {Vianello}, {Weisgarber}, {Westerhoff}, {Wisher}, {Wood},
  {Yapici}, {Yodh}, {Younk}, {Zepeda}, {Zhou}, {Guo}, {Hahn}, {Li}, \&
  {Zhang}}]{Abeysekara2017Sci}
{Abeysekara}, A.~U., {Albert}, A., {Alfaro}, R., {et~al.} 2017{\natexlab{a}},
  Science, 358, 911, \dodoi{10.1126/science.aan4880}

\bibitem[{{Abeysekara} {et~al.}(2017{\natexlab{b}}){Abeysekara}, {Albert},
  {Alfaro}, {Alvarez}, {{\'A}lvarez}, {Arceo}, {Arteaga-Vel{\'a}zquez}, {Ayala
  Solares}, {Barber}, {Bautista-Elivar}, {Becerril}, {Belmont-Moreno},
  {BenZvi}, {Berley}, {Braun}, {Brisbois}, {Caballero-Mora}, {Capistr{\'a}n},
  {Carrami{\~n}ana}, {Casanova}, {Castillo}, {Cotti}, {Cotzomi}, {Couti{\~n}o
  de Le{\'o}n}, {de la Fuente}, {De Le{\'o}n}, {DeYoung}, {Dingus},
  {DuVernois}, {D{\'\i}az-V{\'e}lez}, {Ellsworth}, {Fiorino}, {Fraija},
  {Garc{\'\i}a-Gonz{\'a}lez}, {Gerhardt}, {Gonz{\'a}lez Mun{\"o}z},
  {Gonz{\'a}lez}, {Goodman}, {Hampel-Arias}, {Harding}, {Hernandez},
  {Hernandez-Almada}, {Hinton}, {Hui}, {H{\"u}ntemeyer}, {Iriarte},
  {Jardin-Blicq}, {Joshi}, {Kaufmann}, {Kieda}, {Lara}, {Lauer}, {Lee},
  {Lennarz}, {Le{\'o}n Vargas}, {Linnemann}, {Longinotti}, {Raya},
  {Luna-Garc{\'\i}a}, {L{\'o}pez-Coto}, {Malone}, {Marinelli}, {Martinez},
  {Martinez-Castellanos}, {Mart{\'\i}nez-Castro}, {Mart{\'\i}nez-Huerta},
  {Matthews}, {Miranda-Romagnoli}, {Moreno}, {Mostaf{\'a}}, {Nellen},
  {Newbold}, {Nisa}, {Noriega-Papaqui}, {Pelayo}, {Pretz},
  {P{\'e}rez-P{\'e}rez}, {Ren}, {Rho}, {Rivi{\`e}re}, {Rosa-Gonz{\'a}lez},
  {Rosenberg}, {Ruiz-Velasco}, {Salazar}, {Salesa Greus}, {Sandoval},
  {Schneider}, {Schoorlemmer}, {Sinnis}, {Smith}, {Springer}, {Surajbali},
  {Taboada}, {Tibolla}, {Tollefson}, {Torres}, {Ukwatta}, {Villase{\~n}or},
  {Weisgarber}, {Westerhoff}, {Wisher}, {Wood}, {Yapici}, {Yodh}, {Younk},
  {Zepeda}, \& {Zhou}}]{2017ApJ...843...39A}
---. 2017{\natexlab{b}}, \apj, 843, 39, \dodoi{10.3847/1538-4357/aa7555}

\bibitem[{Cabral \& Leedom(1993)}]{Leedom}
Cabral, B., \& Leedom, L.~C. 1993, in Proceedings of the 20th Annual Conference
  on Computer Graphics and Interactive Techniques, SIGGRAPH '93 (New York, NY,
  USA: Association for Computing Machinery), 263–270.
\newblock \url{https://doi.org/10.1145/166117.166151}

\bibitem[{{Cho}(2019)}]{cho2019ApJ}
{Cho}, J. 2019, \apj, 874, 75, \dodoi{10.3847/1538-4357/ab06f3}

\bibitem[{{Cho} \& {Lazarian}(2003)}]{CL03}
{Cho}, J., \& {Lazarian}, A. 2003, \mnras, 345, 325,
  \dodoi{10.1046/j.1365-8711.2003.06941.x}

\bibitem[{{Cristofari} {et~al.}(2021){Cristofari}, {Blasi}, \&
  {Caprioli}}]{Cristofari2021A&A}
{Cristofari}, P., {Blasi}, P., \& {Caprioli}, D. 2021, \aap, 650, A62,
  \dodoi{10.1051/0004-6361/202140448}

\bibitem[{{Fang} {et~al.}(2019){Fang}, {Bi}, \& {Yin}}]{Fang2019MNRAS}
{Fang}, K., {Bi}, X.-J., \& {Yin}, P.-F. 2019, \mnras, 488, 4074,
  \dodoi{10.1093/mnras/stz1974}

\bibitem[{{Giacinti} \& {Sigl}(2012)}]{Giacinti2012PhRvL}
{Giacinti}, G., \& {Sigl}, G. 2012, \prl, 109, 071101,
  \dodoi{10.1103/PhysRevLett.109.071101}

\bibitem[{{Grebenev} {et~al.}(1995){Grebenev}, {Forman}, {Jones}, \&
  {Murray}}]{Grebenev1995ApJ}
{Grebenev}, S.~A., {Forman}, W., {Jones}, C., \& {Murray}, S. 1995, \apj, 445,
  607, \dodoi{10.1086/175725}

\bibitem[{{Herron} {et~al.}(2018){Herron}, {Burkhart}, {Gaensler}, {Lewis},
  {McClure-Griffiths}, {Bernardi}, {Carretti}, {Haverkorn}, {Kesteven},
  {Poppi}, \& {Staveley-Smith}}]{2018ApJ...855...29H}
{Herron}, C.~A., {Burkhart}, B., {Gaensler}, B.~M., {et~al.} 2018, \apj, 855,
  29, \dodoi{10.3847/1538-4357/aaafd0}

\bibitem[{{Hildebrand} {et~al.}(2009){Hildebrand}, {Kirby}, {Dotson}, {Houde},
  \& {Vaillancourt}}]{hildebrand2009ApJ}
{Hildebrand}, R.~H., {Kirby}, L., {Dotson}, J.~L., {Houde}, M., \&
  {Vaillancourt}, J.~E. 2009, \apj, 696, 567,
  \dodoi{10.1088/0004-637X/696/1/567}

\bibitem[{{Ho} {et~al.}(2019){Ho}, {Yuen}, {Leung}, \&
  {Lazarian}}]{2019ApJ...887..258H}
{Ho}, K.~W., {Yuen}, K.~H., {Leung}, P.~K., \& {Lazarian}, A. 2019, \apj, 887,
  258, \dodoi{10.3847/1538-4357/ab578c}

\bibitem[{{Johnston} {et~al.}(2006){Johnston}, {Karastergiou}, \&
  {Willett}}]{Johnston2006MNRAS}
{Johnston}, S., {Karastergiou}, A., \& {Willett}, K. 2006, \mnras, 369, 1916,
  \dodoi{10.1111/j.1365-2966.2006.10440.x}

\bibitem[{{Johnston} {et~al.}(2007){Johnston}, {Kramer}, {Karastergiou},
  {Hobbs}, {Ord}, \& {Wallman}}]{john2007MNRAS}
{Johnston}, S., {Kramer}, M., {Karastergiou}, A., {et~al.} 2007, \mnras, 381,
  1625, \dodoi{10.1111/j.1365-2966.2007.12352.x}

\bibitem[{{Kandel} {et~al.}(2016){Kandel}, {Lazarian}, \& {Pogosyan}}]{KLP16}
{Kandel}, D., {Lazarian}, A., \& {Pogosyan}, D. 2016, \mnras, 461, 1227,
  \dodoi{10.1093/mnras/stw1296}

\bibitem[{{Lazarian} \& {Pogosyan}(2012)}]{LP12}
{Lazarian}, A., \& {Pogosyan}, D. 2012, \apj, 747, 5,
  \dodoi{10.1088/0004-637X/747/1/5}

\bibitem[{{Lee} {et~al.}(2019){Lee}, {Cho}, \&
  {Lazarian}}]{2019ApJ...877..108L}
{Lee}, H., {Cho}, J., \& {Lazarian}, A. 2019, \apj, 877, 108,
  \dodoi{10.3847/1538-4357/ab1b1e}

\bibitem[{{Lee} {et~al.}(2016){Lee}, {Lazarian}, \&
  {Cho}}]{2016ApJ...831...77L}
{Lee}, H., {Lazarian}, A., \& {Cho}, J. 2016, \apj, 831, 77,
  \dodoi{10.3847/0004-637X/831/1/77}

\bibitem[{Lemoine(2022)}]{Lemoine22}
Lemoine, M. 2022, Phys. Rev. Lett., 129, 215101,
  \dodoi{10.1103/PhysRevLett.129.215101}

\bibitem[{{Liu} {et~al.}(2019{\natexlab{a}}){Liu}, {Ge}, {Sun}, \&
  {Wang}}]{Liu2019ApJ}
{Liu}, R.-Y., {Ge}, C., {Sun}, X.-N., \& {Wang}, X.-Y. 2019{\natexlab{a}},
  \apj, 875, 149, \dodoi{10.3847/1538-4357/ab125c}

\bibitem[{{Liu} {et~al.}(2019{\natexlab{b}}){Liu}, {Yan}, \&
  {Zhang}}]{Liu2019PRL}
{Liu}, R.-Y., {Yan}, H., \& {Zhang}, H. 2019{\natexlab{b}}, \prl, 123, 221103,
  \dodoi{10.1103/PhysRevLett.123.221103}

\bibitem[{{L{\'o}pez-Coto} \& {Giacinti}(2018)}]{lopez2018MNRAS}
{L{\'o}pez-Coto}, R., \& {Giacinti}, G. 2018, \mnras, 479, 4526,
  \dodoi{10.1093/mnras/sty1821}

\bibitem[{{Lorimer} \& {Kramer}(2004)}]{Lorimer2004book}
{Lorimer}, D.~R., \& {Kramer}, M. 2004, {Handbook of Pulsar Astronomy}, Vol.~4

\bibitem[{{Lynn} {et~al.}(2013){Lynn}, {Quataert}, {Chandran}, \&
  {Parrish}}]{Lynn2013}
{Lynn}, J.~W., {Quataert}, E., {Chandran}, B. D.~G., \& {Parrish}, I.~J. 2013,
  \apj, 777, 128, \dodoi{10.1088/0004-637X/777/2/128}

\bibitem[{{Maiti} {et~al.}(2022){Maiti}, {Makwana}, {Zhang}, \&
  {Yan}}]{Maiti2022ApJ}
{Maiti}, S., {Makwana}, K., {Zhang}, H., \& {Yan}, H. 2022, \apj, 926, 94,
  \dodoi{10.3847/1538-4357/ac46c8}

\bibitem[{{Makwana} \& {Yan}(2020)}]{Makwana2020}
{Makwana}, K.~D., \& {Yan}, H. 2020, Physical Review X, 10, 031021,
  \dodoi{10.1103/PhysRevX.10.031021}

\bibitem[{{Malik} {et~al.}(2023){Malik}, {Yuen}, \& {Yan}}]{Malik2023}
{Malik}, S., {Yuen}, K.~H., \& {Yan}, H. 2023, \mnras, 524, 6102,
  \dodoi{10.1093/mnras/stad2225}

\bibitem[{{Martin} {et~al.}(2022){Martin}, {Marcowith}, \&
  {Tibaldo}}]{Martin2022A&A}
{Martin}, P., {Marcowith}, A., \& {Tibaldo}, L. 2022, \aap, 665, A132,
  \dodoi{10.1051/0004-6361/202243481}

\bibitem[{{Pattle} {et~al.}(2017){Pattle}, {Ward-Thompson}, {Berry},
  {Hatchell}, {Chen}, {Pon}, {Koch}, {Kwon}, {Kim}, {Bastien}, {Cho},
  {Coud{\'e}}, {Di Francesco}, {Fuller}, {Furuya}, {Graves}, {Johnstone},
  {Kirk}, {Kwon}, {Lee}, {Matthews}, {Mottram}, {Parsons}, {Sadavoy},
  {Shinnaga}, {Soam}, {Hasegawa}, {Lai}, {Qiu}, \& {Friberg}}]{pattle2017ApJ}
{Pattle}, K., {Ward-Thompson}, D., {Berry}, D., {et~al.} 2017, \apj, 846, 122,
  \dodoi{10.3847/1538-4357/aa80e5}

\bibitem[{{Recchia} {et~al.}(2021){Recchia}, {Di Mauro}, {Aharonian}, {Orusa},
  {Donato}, {Gabici}, \& {Manconi}}]{Recchia2021PhRvD}
{Recchia}, S., {Di Mauro}, M., {Aharonian}, F.~A., {et~al.} 2021, \prd, 104,
  123017, \dodoi{10.1103/PhysRevD.104.123017}

\bibitem[{{Reynolds} {et~al.}(2012){Reynolds}, {Gaensler}, \&
  {Bocchino}}]{Reynolds2012SSR}
{Reynolds}, S.~P., {Gaensler}, B.~M., \& {Bocchino}, F. 2012, \ssr, 166, 231,
  \dodoi{10.1007/s11214-011-9775-y}

\bibitem[{Schlickeiser(2002)}]{schlickeiser2002statistical}
Schlickeiser, R. 2002, Cosmic Ray Astrophysics, 183

\bibitem[{{Sofue} {et~al.}(2019){Sofue}, {Nakanishi}, \&
  {Ichiki}}]{Sofue2019MNRAS}
{Sofue}, Y., {Nakanishi}, H., \& {Ichiki}, K. 2019, \mnras, 485, 924,
  \dodoi{10.1093/mnras/stz407}

\bibitem[{{Takahashi}(2023)}]{Takahashi2023PASJ}
{Takahashi}, K. 2023, \pasj, 75, S50, \dodoi{10.1093/pasj/psac111}

\bibitem[{{Uyaniker} {et~al.}(1999){Uyaniker}, {F{\"u}rst}, {Reich}, {Reich},
  \& {Wielebinski}}]{Uyaniker1999A}
{Uyaniker}, B., {F{\"u}rst}, E., {Reich}, W., {Reich}, P., \& {Wielebinski}, R.
  1999, \aaps, 138, 31, \dodoi{10.1051/aas:1999494}

\bibitem[{{Wang} {et~al.}(2020){Wang}, {Zhang}, \& {Xiang}}]{wang2020}
{Wang}, R.-Y., {Zhang}, J.-F., \& {Xiang}, F.-Y. 2020, \apj, 890, 70,
  \dodoi{10.3847/1538-4357/ab6a1a}

\bibitem[{{Yan}(2022)}]{Yan2022rev}
{Yan}, H. 2022, in 37th International Cosmic Ray Conference, 38

\bibitem[{{Yan} \& {Lazarian}(2002)}]{2002PhRvL..89B1102Y}
{Yan}, H., \& {Lazarian}, A. 2002, \prl, 89, 281102,
  \dodoi{10.1103/PhysRevLett.89.281102}

\bibitem[{{Yan} \& {Lazarian}(2004)}]{YL2004ApJ}
---. 2004, \apj, 614, 757, \dodoi{10.1086/423733}

\bibitem[{{Yan} \& {Lazarian}(2008)}]{Yan2008ApJ}
---. 2008, \apj, 673, 942, \dodoi{10.1086/524771}

\bibitem[{{Yan} {et~al.}(2008){Yan}, {Lazarian}, \& {Petrosian}}]{YLP2008}
{Yan}, H., {Lazarian}, A., \& {Petrosian}, V. 2008, \apj, 684, 1461,
  \dodoi{10.1086/589962}

\bibitem[{{Yan} {et~al.}(2012){Yan}, {Lazarian}, \&
  {Schlickeiser}}]{Yan2012ApJCR}
{Yan}, H., {Lazarian}, A., \& {Schlickeiser}, R. 2012, \apj, 745, 140,
  \dodoi{10.1088/0004-637X/745/2/140}

\bibitem[{{Yuen} {et~al.}(2023){Yuen}, {Yan}, \& {Lazarian}}]{leakage}
{Yuen}, K.~H., {Yan}, H., \& {Lazarian}, A. 2023, \mnras, 521, 530,
  \dodoi{10.1093/mnras/stad287}

\bibitem[{Zöckler {et~al.}(1997)Zöckler, Stalling, \& Hege}]{ZOCKLER1997975}
Zöckler, M., Stalling, D., \& Hege, H.-C. 1997, Parallel Computing, 23, 975,
  \dodoi{https://doi.org/10.1016/S0167-8191(97)00039-2}

\end{thebibliography}
\label{lastpage}
\end{document}